\documentclass[preprint,10pt]{aastex}
\usepackage{graphicx}

\slugcomment{\today}

\begin{document} 
\title{Determining Foreground Contamination in CMB Observations:\\
Diffuse Galactic Emission in the MAXIMA-I Field}

\author{
A.~H.~Jaffe\altaffilmark{1},
A.~Balbi\altaffilmark{2}, 
J.~R.~Bond\altaffilmark{3}, 
J.~Borrill\altaffilmark{4}, 
P.~G.~Ferreira\altaffilmark{5},
D.~Finkbeiner\altaffilmark{6},
S.~Hanany\altaffilmark{7}, 
A.~T.~Lee\altaffilmark{8,9},
B.~Rabii\altaffilmark{8},
P.~L.~Richards\altaffilmark{8},
G.~F.~Smoot\altaffilmark{8,10},
R.~Stompor\altaffilmark{4}, 
C.~D.~Winant\altaffilmark{8},
J.~H.~P. ~Wu\altaffilmark{11} }

\altaffiltext{1}{Astrophysics Group, Blackett Lab, Imperial College,
  London SW7 2BW, UK}
\altaffiltext{2}{Dipartimento di Fisica, Universit\`a Tor Vergata
  Roma, Italy}
\altaffiltext{3}{CITA, University of Toronto, 60 St George St., Toronto,
  Canada}
\altaffiltext{4}{National Energy Research
Scientific Computational Research Division, Lawrence Berkeley National Laboratory,
Berkeley, CA, USA}
\altaffiltext{5}{Astrophysics, University of Oxford, Oxford, UK}
\altaffiltext{6}{Hubble Fellow; Princeton University, Department of
  Astrophysics, Peyton Hall, Princeton, NJ 08544} 
\altaffiltext{7} {School of Physics and Astronomy, 
University of Minnesota/Twin Cities, Minneapolis, MN, USA}
\altaffiltext{8}{Dept. of Physics, University of California, Berkeley
CA, USA}
\altaffiltext{9}{Space Sciences Laboratory, University of California,
  Berkeley, CA, USA}
\altaffiltext{10}{Division of Physics,Lawrence Berkeley National Laboratory, 
Berkeley, CA, USA}
\altaffiltext{11}{National Taiwanese University, Taipei, Taiwan}


\begin{abstract}
  
  Observations of the CMB can be contaminated by diffuse foreground
  emission from sources such as Galactic dust and synchrotron radiation.
  In these cases, the morphology of the contaminating source is known
  from observations at different frequencies, but not its amplitude at
  the frequency of interest for the CMB.  We develop a technique for
  accounting for the effects of such emission in this case, and for
  simultaneously estimating the foreground amplitude in the CMB
  observations. We apply the technique to CMB data from the MAXIMA-1
  experiment, using maps of Galactic dust emission from combinations of
  IRAS and DIRBE observations, as well as compilations of Galactic
  synchrotron emission observations. The spectrum of the dust emission
  over the 150--450 GHz observed by MAXIMA is consistent with preferred
  models but the effect on CMB power spectrum observations is
  negligible.

\end{abstract}


\section{Introduction} 

Measurements of the Cosmic Microwave Background (CMB) have begun to
fulfill their promise to image the Universe at the epoch of the
decoupling of photons from baryons, thereby measuring cosmological
parameters to new levels of precision. In recent years, the
balloon-borne bolometer experiments MAXIMA-1 \citep{Hanany00,Lee01,Rabii03instrument},
BOOMERANG \citep{debern00,Nett01} and Archeops
\citep{BenoitArcheops2002}, as well as the ground-based interferometers,
CBI \citep{CBI,CBIDeep}, DASI \citep{DASI2} and VSA
\citep{GraingeVSA2002} have measured the CMB power spectrum down to
angular scales of 10 arcminutes or better; most recently, WMAP has
observed the full CMB sky over a factor of four in frequency and with a
beam of 13 arcmin FWHM. The CMB photons are produced at the last
scattering surface, when the opaque, charged plasma of electrons and
nuclei becomes a transparent gas of hydrogen and helium, at a
temperature of about 1~eV, the epoch of ``Recombination.''  The measured
signals thus map the primordial distribution of matter in the Universe,
and allow measurements of the cosmological parameters and an
understanding of the mechanism for laying down the initial fluctuations.

This promise has, however, always been tempered by the possibility of
contamination from astrophysical sources of microwave emission or
absorption, even away from the Galactic plane. Happily, the cosmological
signal has proven to be dominant over a wide range of frequencies and
angular scales. Nonetheless, we can hope to isolate these foreground
contributions: the CMB itself has the well-understood shape of a
black-body spectrum at a given temperature. Other contributions,
Galactic or extragalactic, will in all but the most pathological cases
have a different spectral dependence. While the peak of the 2.73K CMB
intensity lies at about 90 GHz, other contributions dominate at other
frequencies \citep[e.g.,][]{TegEisHuDOC00}. Emission from dust (and
dusty external galaxies) is expected to dominate the CMB at higher
frequencies; synchrotron and free-free emission (and galaxies radiating
by these methods) dominate at lower frequencies.

Very often, then, we will have independent measurements of the
contribution from a given foreground component. However, because those
measurements are taken at a different frequency than those of the CMB
measurements, we must extrapolate down to the frequencies of interest.
As we inevitably lack a perfect understanding of the foreground emission
mechanism, this extrapolation will be imprecise.

In this work, we present a framework for dealing with such imprecise
extrapolation, by leaving the spectral dependence free to vary, but
fixing the morphology to be determined by the external measurements. The
method allows us, on the one hand, to determine the global spectral behavior of
the foreground component and ``marginalize over'' (in Bayesian parlance)
the CMB signal, or, on the other hand, to determine the CMB signal while
marginalizing over the foreground contribution.  Similar methods have
been proposed in the past with more ad hoc derivations \citep[e.g.,][and
references
therein]{DodStebFG94,TegEfsFG96,DodFG97,TegFG98,TegCostaProc}. Here, we
show how the same formalism can deal with several different foreground
problems: estimating the amplitude of the foreground emission;
estimating the CMB map after accounting for the foregrounds; and finally
estimating the CMB power spectrum in the presence of such contamination.

In this paper we concentrate on the contribution of Galactic dust
emission to the emission observed by the MAXIMA-1 experiment at 150--410
GHz. As our foreground template, we use the recent seminal work of
\citet{FinDavSch99}, who combined data from the IRAS satellite with that from
the DIRBE and FIRAS instruments on the COBE satellite to extrapolate the
spectrum of Galactic dust pixel-by-pixel to our CMB frequencies.  These
maps were also used in the analysis of the dust signal present in
BOOMERANG data \citep{Masi01}.

Finally, we note that the recent work of the WMAP team takes a somewhat
different approach, using a maximum-entropy method to estimate
foreground emission \citep{WMAPforegrounds}. In that work, they too use
the \citet{FinDavSch99} maps as a ``prior'' for the dust emission. 
However, the effective sensitivity of the WMAP dust reconstruction is
comparable to the CMB sensitivity.
Thus, in regions of very low dust contrast such as that observed by
MAXIMA-1, the WMAP prediction of dust emission is insufficiently
sensitive, and indeed imperfectly correlated with the input prior
maps.

\section{Methods}
\label{sec:methods}

In this paper, we are concerned with the possible contribution from the
aforementioned sources of Galactic foreground emission, in particular
contamination by dust emission. We will further assume that we have a
robust measurement of the foreground at some frequency other than that
with which we observe the CMB. For example, dust is dominant at high
frequencies, $f\gtrsim300$ GHz, but believed to be only a small
contaminant at the 50--200 GHz at which the CMB dominates high-Galactic
latitude emission.

Thus, we will assume that the spatial pattern of these sources is known,
and described by some template, $f_{ip}$, where $p$ numbers the observed
pixels in our CMB maps, and $i$ labels the various foreground
contributions (e.g., dust, synchrotron, etc.).  However, we will leave
the {\em amplitude} of the individual foreground contributions to be
determined. That is, $f_{ip}$ is the foreground map, at its observed
frequency for example. This will be multiplied by an unknown amplitude
$\beta_i$, to be determined as the result of our analysis.

For the purposes of this paper, then, the intensity in a pixel of our
CMB map is given by
\begin{equation}
  \label{eq:data}
  d_p = \sum_i\beta_i f_{ip} + s_p  + n_p
\end{equation}
where $d_p$ is the data at pixel $p$; $s_p$ is the CMB signal at that
pixel; $n_p$ is the noise in the pixel; and $\sum_i\beta_i f_{ip}$
gives the total contribution from the foreground sources under
consideration.  We take the CMB signal and the foreground template to be
already smeared by the beam and any other instrumental effects:
\begin{eqnarray}
  \label{eq:sp}
  s_p &=& \int d^2{\bf\hat x}\; B({\bf\hat x},{\bf\hat x}_p) T({\bf\hat x})
      \nonumber\\
      &=& \int d^2{\bf\hat x} \;B({\bf\hat x}\cdot{\bf\hat x}_p)
      T({\bf\hat x})\nonumber\\
      &=& \sum_{\ell m} B_\ell a_{\ell m} Y_{\ell m}({\bf\hat x}_p)\;,
\end{eqnarray}
where $B({\bf\hat x},{\bf\hat x}_p)$ gives the response of the beam at
position ${\bf\hat x}$ when pointed at pixel $p$. In the second equality
we assume that the beam is azimuthally symmetric around pixel $p$, and
in the third equality we transform to spherical harmonics with indices
$(\ell,m)$. Here, $B_\ell$ is the transform of the symmetric
$B({\bf\hat x}\cdot{\bf\hat x}_p)$ and $a_{\ell m}$ is the transform of
$T({\bf\hat x})$.  For more information on working with asymmetric beams
and pixels, see \citet{Wu01beams,SouradeepBeams}.

Note that this model, Eq.~\ref{eq:data}, is essentially the same as
that considered in \citet{Stompor02}, although that work considers a
broader class of templates, $f_{ip}$, allowing them to be associated
with an amplitude synchronous with 
instrumental characteristics rather than the sky. That work then derives
techniques, analogous to the ones described below, to determine or
marginalize over these amplitudes.

Under this model, we can ask two separate questions: 
\begin{enumerate}
\item What are the foreground amplitudes, $\beta_i$?
\item What is the underlying signal, $s_p$? Perhaps more important is the
  related question: What is $C_\ell$? That is, what are the statistics
  of the cosmological component, $s_p$, taking into account the
  possible presence of foreground contamination?
\end{enumerate}

To answer these questions we must first assign appropriate distributions
to the quantities in Eq.~\ref{eq:data}. As usual, the noise is taken to
be distributed as a zero-mean Gaussian with covariance $N_{pp'} \equiv
\langle n_p n_{p'} \rangle$ (assumed known beforehand, although we can
also apply iterative techniques \citep{FerJafMNRAS00,Dore01} to determine it
simultaneously with the CMB and foreground signals). This gives a
likelihood function
\begin{eqnarray}
  \label{eq:like1}
  P(d|\beta, s) &=& \frac{1}{\left|2\pi N\right|^{1/2}} \times\nonumber\\
  &&\exp\left[{-\frac12 (d-s-\beta f)^T N^{-1} (d-s-\beta f)}\right]\; ,
\end{eqnarray}
where we have left off indices and used matrix notation.


In order to determine the underlying signal power spectrum, we will need to
assign an appropriate Gaussian distribution, with variance given by
\begin{equation}
  \label{eq:Spp}
  S_{pp'}\equiv \langle s_p s_{p'} \rangle
= \sum_\ell \frac{2\ell+1}{4\pi} B^2_\ell C_\ell P_\ell(\cos\theta_{pp'}),
\end{equation}
where $C_\ell\equiv\langle|a_{\ell m}|^2\rangle$ is the
underlying CMB power spectrum, $\theta_{pp'}$ is the angle between
pixels $p$ and $p'$, and the $P_\ell$ are the Legendre polynomials. This
gives a distribution
\begin{equation}
  \label{eq:sPrior}
  P(s|C_\ell) =  \frac{1}{\left|2\pi N\right|^{1/2}} 
  \exp\left[-\frac12 s^T S^{-1} s\right]\;.
\end{equation}

All of this is as in the usual foreground-free case.  The likelihood for
the data $d$ (given $\beta$ and $s$) is a Gaussian, now with mean
$\sum_i \beta_i f_{ip} + s_p$. Finally, however, we must assign a prior
to the amplitudes $\beta_i$. To remain sufficiently general, we will
allow an arbitrary Gaussian for each amplitude, with\footnote{If
  desired, we can include a known mean for the foreground contribution
  by simply subtracting it from the data at the outset.}
$\langle\beta_i\rangle = 0$, and $\langle \beta^2 \rangle =
\sigma_{\beta i}^2$. The prior is then
\begin{equation}
  \label{eq:betaPrior}
  P(\beta) = \prod_i \frac{1}{\sqrt{2\pi\sigma_i^2}}  
  \exp\left[-\frac12 \frac{\beta_i^2}{\sigma_{\beta i}^2}\right]\;.
\end{equation}
We can take $\sigma_{\beta i} \to \infty$ to give a `non-informative'
distribution. This is algebraically easier to deal with than the
equivalent unbounded uniform distribution.

If we combine these priors with the likelihood, we use Bayes' theorem to
get the posterior distribution
\begin{equation}\label{eq:post}
  P(\beta_i, s_p | C_\ell, d_p) \propto P(\beta_i) P(s_p|C_\ell)
  P(d_p | \beta_i , s_p, C_\ell) 
\end{equation}
where $P(\beta_i)$ and $P(s_p|C_\ell)$ are priors, and
$P(d_p | \beta_i , s_p)$ is the
likelihood.  Then, in order to answer each of the above questions, we
marginalize over $\beta$ to get the posterior for $s$; $s$ to get the
posterior for $\beta$; and finally marginalize over both, after giving
$s$ the appropriate prior variance appropriate for a given $C_\ell$.
Because of the linear, Gaussian form of our likelihood and priors, all
of these marginalizations essentially involve ``completing the square,''
and each of the posteriors remains a Gaussian distribution. For the same
reason, these results can also be derived on the usual minimum-variance
grounds rather than in this Bayesian formalism.

First, we calculate the posterior distribution of $\beta$ by 
marginalizing over $s_p$:
\begin{eqnarray}
  \label{eq:betapost}
  P(\beta  | d_p, C_\ell) &=& \int ds\; P(\beta, s | d, C_\ell) \nonumber\\
&=&\frac{1}{|2\pi M_\beta |^{1/2}}\times\nonumber\\
&&\exp\left[
  -\frac12(\beta-{\bar\beta})^T M_\beta^{-1}(\beta-{\bar\beta}) \right]
\end{eqnarray}
This is a Gaussian with mean
\begin{equation}
  \label{eq:betamean}
  {\bar\beta}_i =  \langle\beta_i\rangle = 
  \left[f^T (S + N)^{-1} f\right]_{ii'}^{-1} (f^T (S + N)^{-1} d)_{i'}
\end{equation}
and covariance 
\begin{equation}
  \label{eq:betavar}
M_{\beta,ii'}=
  \langle (\beta - {\bar\beta}_i) (\beta - {\bar\beta}_{i'})\rangle = 
\left[f^T (S + N)^{-1} f\right]_{ii'}^{-1}\;.
\end{equation}
We have already taken the prior variances $\sigma_{\beta i}\to\infty$ in this
expression. As is often the case in linear problems with normal errors
as we have here, this is just the least squares solution for the
amplitudes, $\beta_i$.


The posterior for the signal, $s_p$, is also a Gaussian. It has mean
\begin{equation}
\label{eq:wiener}
{\bar s}_p = S(S+N+\sigma_\beta^2 f f^T)^{-1} d
\end{equation}
and covariance
\begin{eqnarray}
  \langle (s - {\bar s}_p) (s - {\bar s}_{p'})\rangle &=& 
S (S+N+\sigma_\beta^2 f f^T)^{-1} (N+\sigma_\beta^2 f f^T)\nonumber\\
 &=& S-S(S+N+\sigma_\beta^2 f f^T)^{-1}S\;.
\end{eqnarray}
This gives us the \emph{Wiener Filter} as the mean of this distribution.
Note that we have left in a finite prior variance for $\beta_i$.  We
can take these prior variances to infinity using the
Sherman-Morrison-Woodbury formula, which states
\begin{equation}
  \label{eq:ShermanMorrison}
  (W+f B f^T)^{-1} = W^{-1} - W^{-1} f (f^T W^{-1} f -
  B^{-1})^{-1} f^T W^{-1} \; .
\end{equation}
Setting $W=S+N$ and $B^{-1}=\sigma_\beta^{-2} I\to0$ here is equivalent
to marginalizing over our template amplitudes.
Note that, as in \citet{Stompor02}, this same formula can be applied in
the determination of the original map, $d_p$, in order to marginalize
over such modes at an earlier stage.

Finally, we wish to determine the power spectrum, $C_\ell$. We do this
by starting with the expression in Eq.~\ref{eq:post}, and using Bayes'
theorem yet again:
\begin{equation}
  \label{eq:PCl}
  P(C_\ell|d_p) \propto P(C_\ell) P(d_p|C_\ell)
 = P(C_\ell) \int d\beta \; ds P(\beta,s|d,C_\ell)\;.
\end{equation}
(Equivalently, we could have considered $C_\ell$ a ``parameter'' from the
start, and just given it a delta function prior when calculating the
distributions of $\beta$ and $s$.)  In this case, it is worth writing
out the entire expression:
\begin{equation}
  \label{eq:Cllike}
  P(C_\ell | d ) = P(C_\ell) \frac{1}{\left|2\pi M\right|^{1/2}} 
  \exp\left[-\frac12 d^T M^{-1} d \right]
\end{equation}
where the covariance matrix is given by
\begin{eqnarray}
  \label{eq:dmatrix}
 M_{pp'} &=&   \langle d_p d_{p'} \rangle =
  \sum_{ii'} f_{ip} \langle\beta_i\beta_{i'}\rangle f_{i'p'} + 
\langle s_p s_{p'} \rangle + \langle n_p n_{p'} \rangle \nonumber\\
&=& (\sigma_\beta^2 f^T f)_{pp'} + S_{pp'} + N_{pp'}\;.
\end{eqnarray}
Recall that the signal covariance is a function of the underlying power
spectrum, $C_\ell$, as in Eq.~\ref{eq:Spp}.  We then maximize this with
respect to the $C_\ell$ (or bands thereof with known shape) as in
\citet{BJK98} as implemented in
MADCAP\footnote{\url{http://www.nersc.gov/$\sim$borrill/cmb/madcap.html}}
\citep{MADCAP}. In this expression, we have kept a finite variance for
the dust prior, $\sigma^2_\beta$. In the absence of a known dust
contaminant spectrum (or to be most conservative), we can take
$\sigma_\beta\to\infty$.  The numerical implementation of this can be
done using the Sherman-Morrison-Woodbury formula, implemented in the
current version of MADCAP.

Note that it is the linear nature of our model Eq.~\ref{eq:data}, along
with our assignment of Gaussian priors and likelihoods, that allows
these analytic simplifications. This makes it somewhat more difficult to
determine or marginalize over more physical parameters such as, say, the
spectral index of the foreground spectra. However, we can use this
formalism to determine several $\beta_i$, each corresponding to a
different region of the sky --- allowing us to take account at
some level of spatial variations in the spectrum. For the present case,
however, the final signal-to-noise is too low for this to be fruitful.

In the Bayesian picture, our method simply ignores any information
associated with a pattern on the sky matching our chosen templates.
Thus, even if our template were ``wrong'' (i.e., if they did not
accurately reflect the pattern of foreground emission on the sky), the
final estimate of the power spectrum, including its error bars, would be
no less correct than that without applying this method at all. In the
simplest case, the maximum likelihood power could decrease, but the
error bars would increase to take this into account.

\subsection{Noisy Foreground Templates}

So far, we have only considered foreground templates, $f$, that are
accurately known. In many cases, however, the pattern will also be the
result of a noisy measurement. That is, we supplement our model of the
data, Eq.~\ref{eq:data} with a new set of pixel measurements,
\begin{equation}
  \label{eq:fdata}
  e_p = f_p + \nu_p
\end{equation}
where $e_p$ is the new data, $\nu_p$ is noise, satisfying
$\langle\nu_p\nu_{p'}\rangle=E_{pp'}$, and we can again assign a
Gaussian error distribution for $\nu_p$. Now, to proceed we must
multiply this new distribution by the likelihood of
Eq.~\ref{eq:like1} or the full Eq.~\ref{eq:post}, and marginalize over
the now unknown template, $f$:
\begin{eqnarray}
  \label{eq:flike}
  P(de|\beta, s) &=& \frac{1}{\left|2\pi N + \beta^2 E\right|^{1/2}} \times\nonumber\\
  &&\exp\left[{-\frac12 (d-s-\beta e)^T (N+\beta^2E)^{-1} (d-s-\beta e)}\right]\; .
\end{eqnarray}
After marginalization, the variance is increased by $\beta^2E$ over the
case where the foreground template is known exactly. As before, we can
also put in a prior for the signal amplitude, $s$ (Eq.~\ref{eq:sPrior}),
and marginalize, giving the equivalent of Eq.~\ref{eq:Cllike},
\begin{equation}
  \label{eq:fCllike}
  P(C_\ell | de ) = P(C_\ell) \frac{1}{\left|2\pi (S+N+\beta^2E)\right|^{1/2}} 
  \exp\left[-\frac12 (d-\beta e)^T (S+N+\beta^2E)^{-1} (d-\beta e) \right]\;.
\end{equation}
In both cases, the marginalization over the foreground template has
destroyed the linear nature of the problem, and the unknown amplitude,
$\beta$ now appears in the effective correlation matrix; we can no
longer find analytic formulae to determine nor marginalize over the amplitude

In fact, the problem for a noisy template is formally identical to the
cross-calibration of cross-comparison problem, as discussed in
\cite{Knoxetal1998}: the amplitude, $\beta$ is just the relative
calibration of the foreground contribution to the two datasets. As in
that work, however, we must use somewhat more laborious numerical
techniques to deal with the more complicated problem.

\section{Application: Diffuse foregrounds and MAXIMA-1}

Here we apply these techniques to the data from the MAXIMA-I
experiment; the hardware is described in \citet{Lee99} the resulting
maps and power spectra are described in \citet{Hanany00,Lee01}, the cosmological
implications are presented in \citet{Balbi00,Stompor01,maxiboom01} and details of the
data analysis in \citet{Stompor02}. In this work, we consider maps made
from the data of individual photometers at various frequencies: three at
150 GHz, two at 240 GHz, and two at 410 GHz. Note that only a subset of
these maps were used for the cosmological results presented in the
previous MAXIMA-I papers. A full account of the systematic checks on the
MAXIMA-I data and on the consistency between channels is given in
\citet{Stompor03systematics}.

The primary aim of the MAXIMA experiment is to observe the cosmological
component of the CMB. The MAXIMA-I field was thus explicitly chosen to be
low in Galactic foreground emission. Nonetheless, our methods let us
take advantage of the large number of pixels simultaneously, even though
the dust contribution is insignificant in a single one. 

\subsection{Dust Emission from Galactic Cirrus}

As foreground templates, we primarily work with the combined IRAS/DIRBE
infrared dust maps presented in \citet{SchFinDav98}.  These are
extrapolated from 100-240\micron\ to our CMB observation frequencies
using dust emissivity models constrained by COBE/FIRAS spectral
observations \citep{FinDavSch99}. One can use these emissivity models to
compute a dust temperature and column density in each pixel from the
DIRBE 100 and 240\micron\ maps, and extrapolate to much lower
frequencies.  Although our formalism could be used to extrapolate from
the observed infrared frequencies down to our observations at 100-400
GHz, neglect of the dust temperature variation from pixel to pixel can
cause errors of a factor of $\sim 2$, so we use the FDS99 models that
already take account of this variation explicitly.  Eight models are
computed; four single-model components (power laws) and four
two-component models (where the two components have a fixed mass ratio,
and have temperature coupled to the radiation field in a self-consistent
way).  The parameters of the model are given in Table 1, adapted from
FDS99.  The 4 parameters describing each model are: $\alpha_1,
\alpha_2$, the emissivity power-law indices for components one and two,
$f_1$, the fraction of power absorbed and re-emitted by component 1
(unrelated to the foreground template vector $f$ above), and $q_1/q_2$
where $q_i$ is the IR/optical opacity ratio for component $i$.  The
two-component models listed in the table are: 5) $T_1=T_2$, indices from
\citet{Pollack94}, 6) $\alpha_1=\alpha_2=2$ as in \citet{Reach95}, 7)
like 5 except temperatures can float, and 8) all four parameters
floating.  Because 8 gives the best chi-squared, it is the preferred
model, although 7) is not appreciably different. Here, we will
concentrate on their overall best-fit model 8, with $\alpha_1=1.67$,
$\alpha_2=2.70$, $f_1=0.0363$, and $q_1/q_2=13.0$.

\begin{figure*}[htbp]
  \begin{center}
  \includegraphics[width=0.4\columnwidth,angle=90]{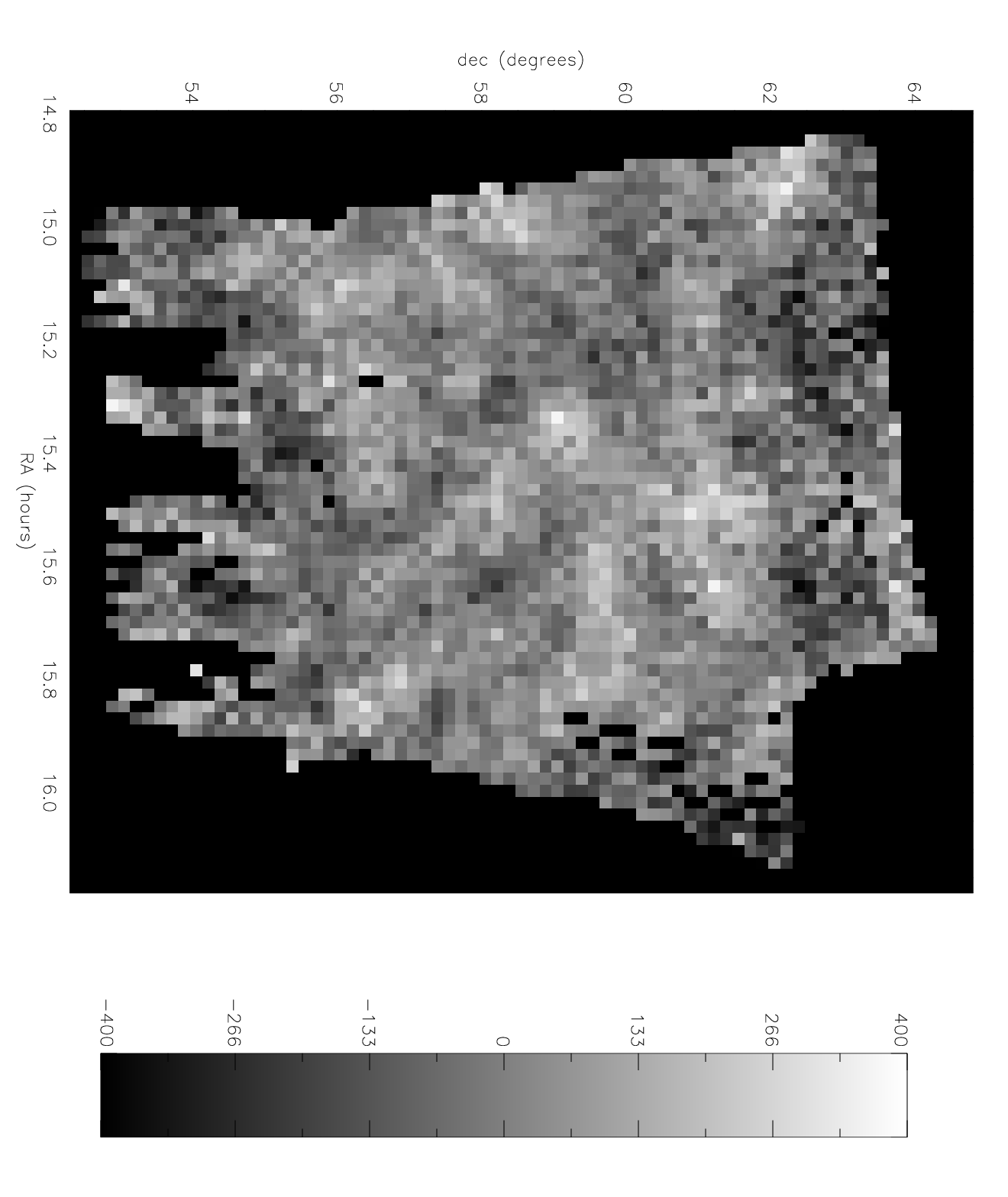}
  \includegraphics[width=0.4\columnwidth,angle=90]{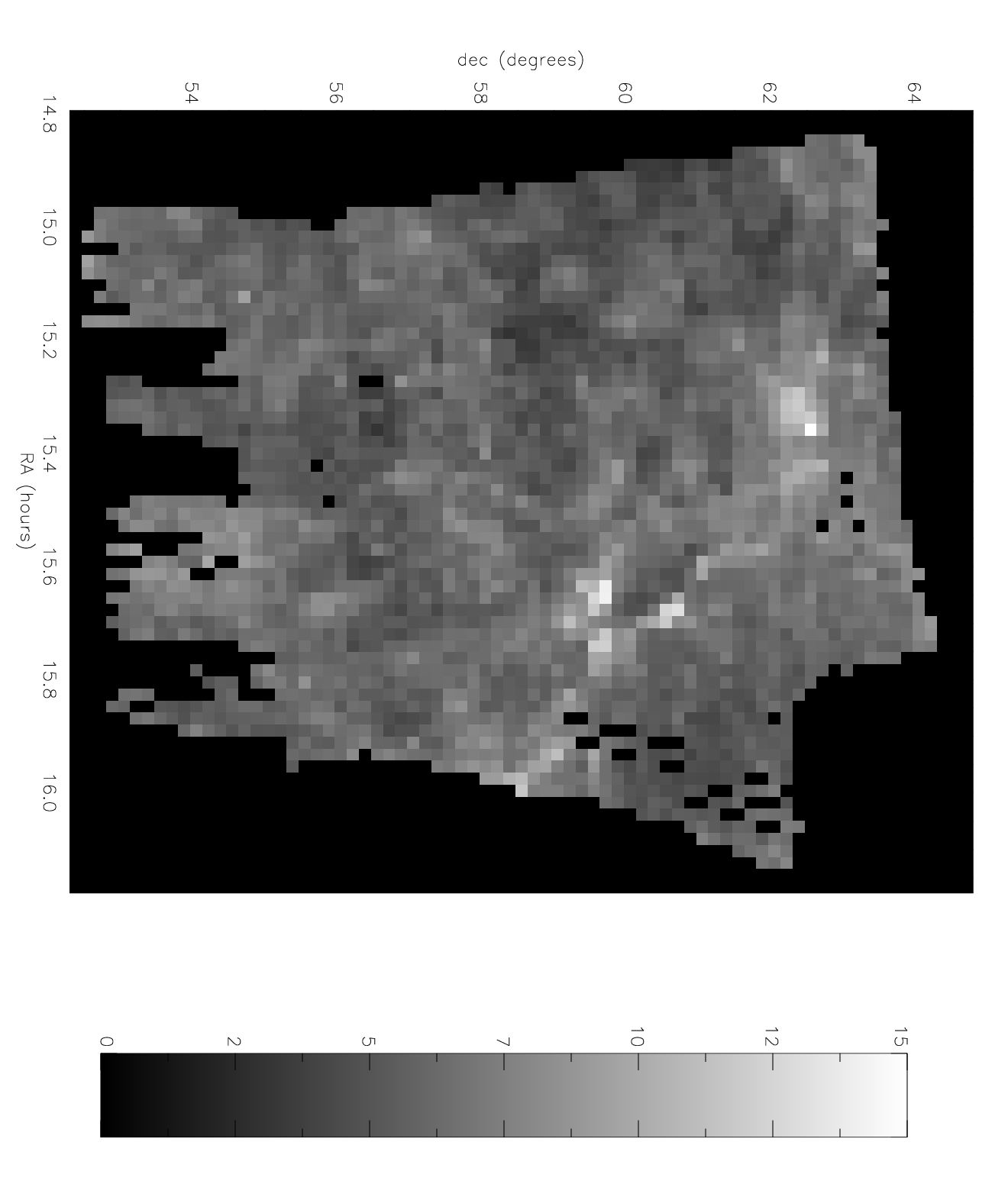}
    \caption{Left: MAXIMA-I map of microwave emission at 150 Ghz. Right:
      Map of dust emission, extrapolated from IRAS/DIRBE maps to 150
      GHz. Colorbars give the temperature in $\mu$K. Note that the difference
      in the temperature range between the total and dust emission is a
      factor of $\sim50$.}
    \label{fig:dustmaps150}
  \end{center}
\end{figure*}

\begin{figure*}[htbp]
  \begin{center}
  \includegraphics[width=0.4\columnwidth,angle=90]{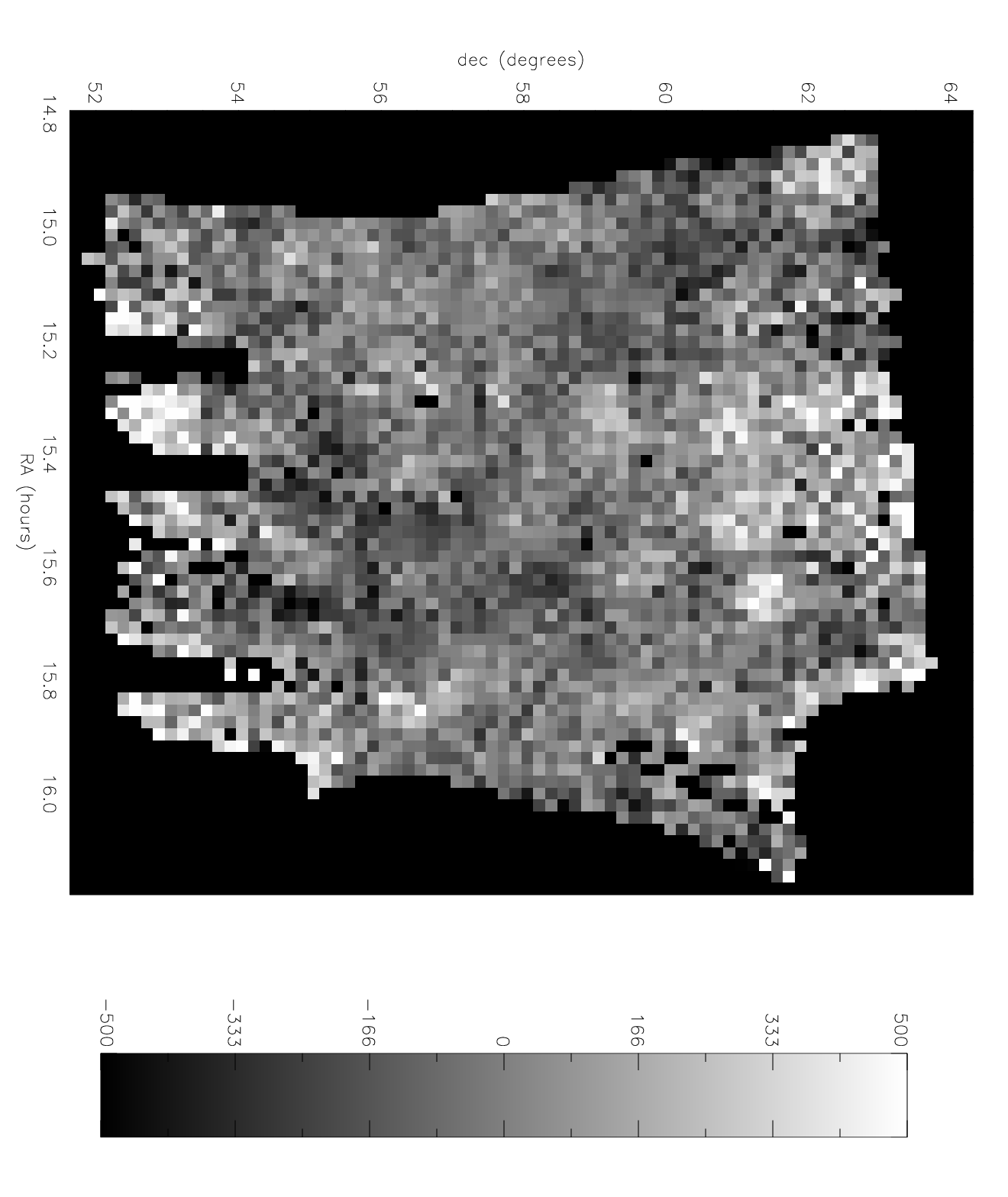}
  \includegraphics[width=0.4\columnwidth,angle=90]{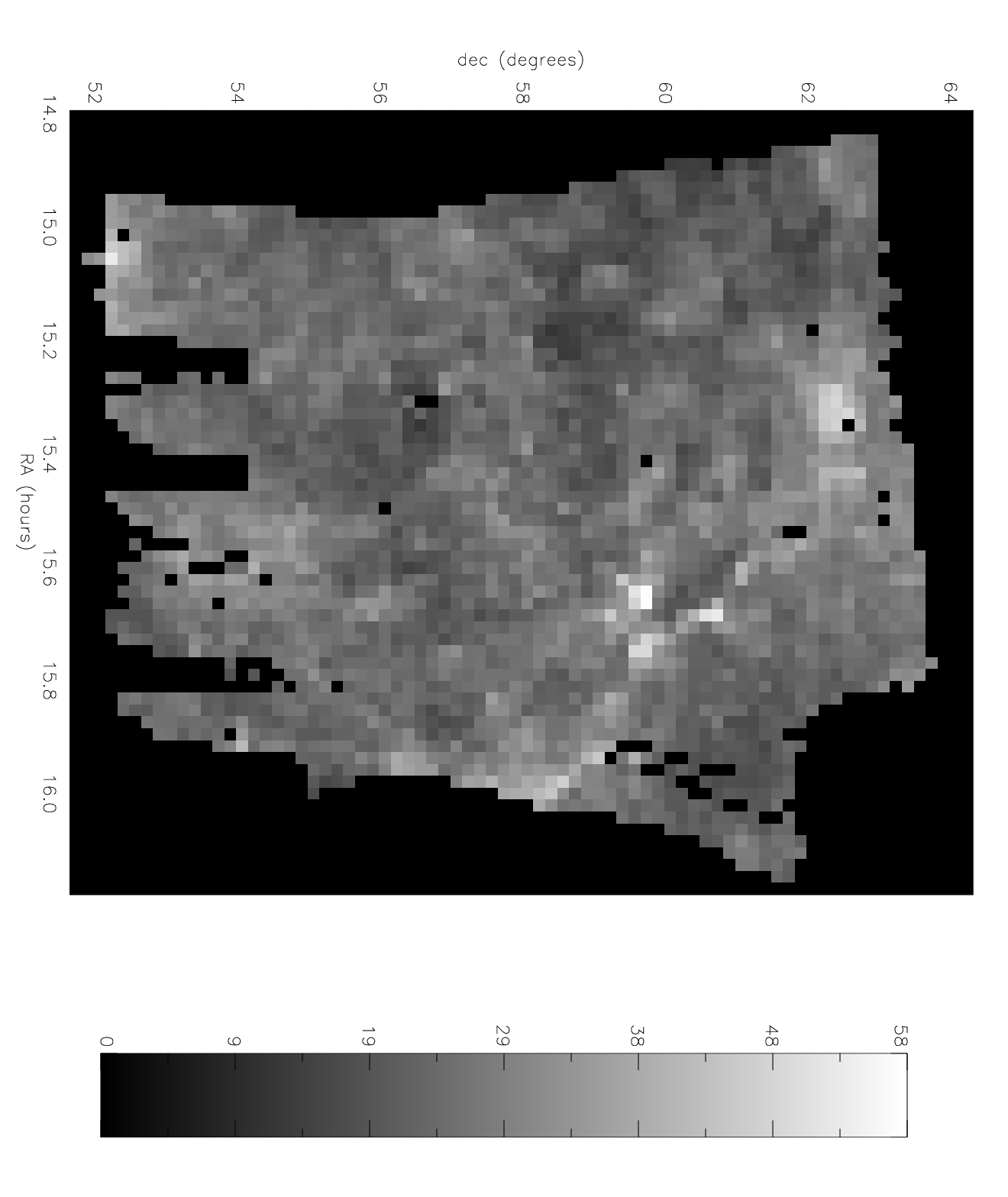}
    \caption{Left: MAXIMA-I map of CMB Emission at 240 Ghz. Right:
      Map of dust emission, extrapolated from IRAS/DIRBE maps to 240
      GHz. Note that the difference
      in the temperature range between the total and dust emission is a
      factor of $\sim17$. }
    \label{fig:dustmaps240}
  \end{center}
\end{figure*}

\begin{figure*}[htbp]
  \begin{center}
  \includegraphics[width=0.4\columnwidth,angle=90]{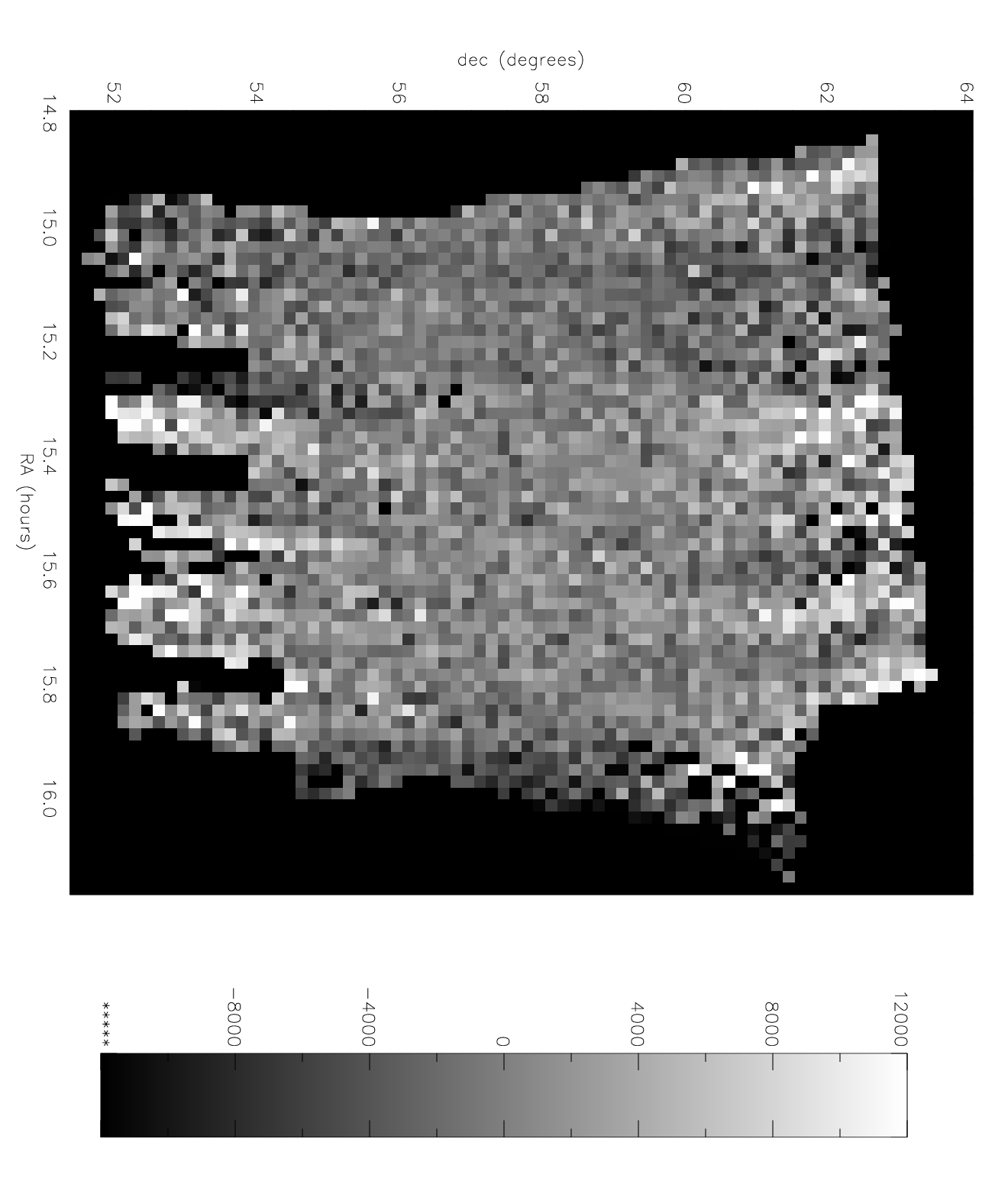}
  \includegraphics[width=0.4\columnwidth,angle=90]{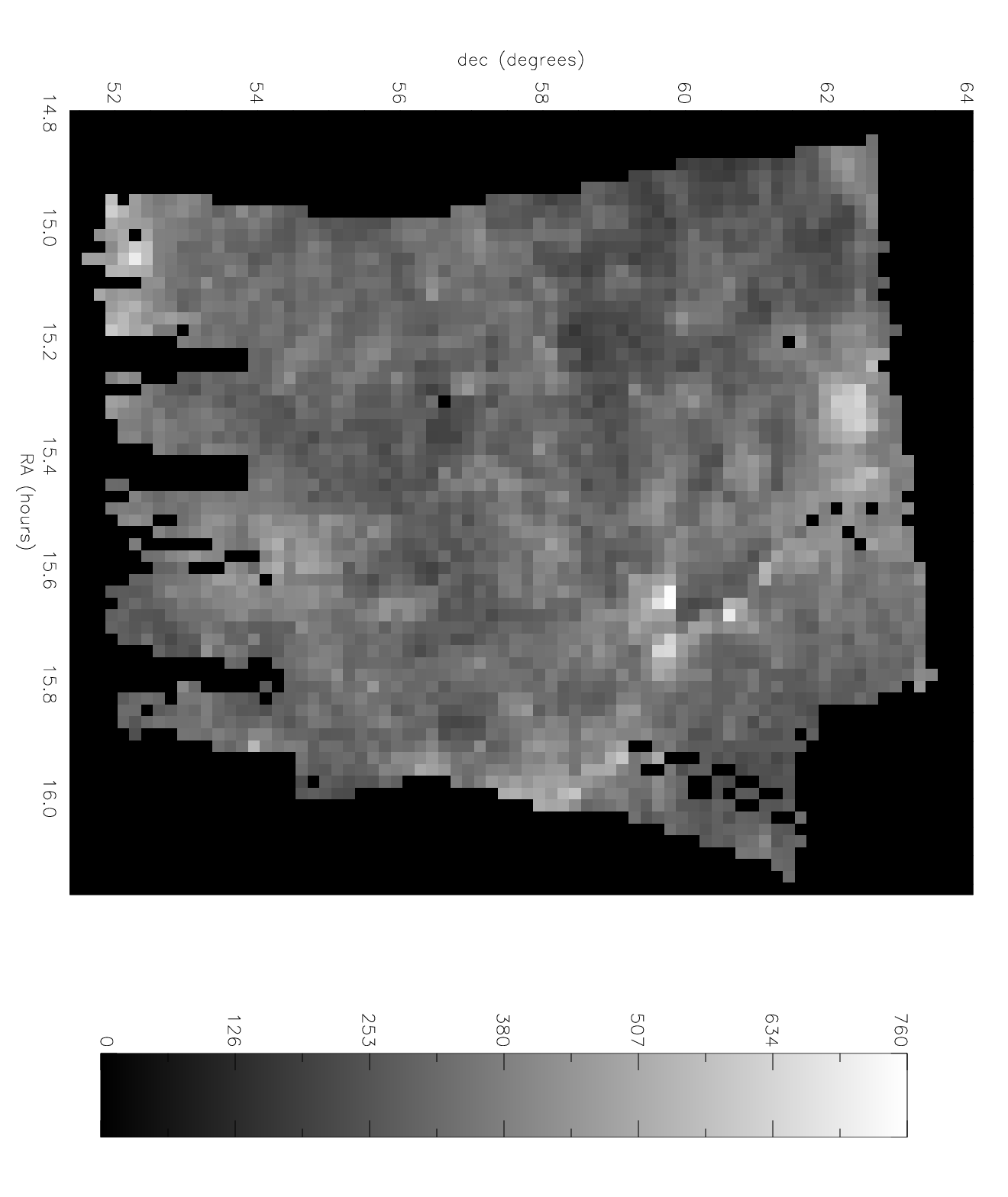}
    \caption{Left: MAXIMA-I map of CMB Emission at 410 Ghz. Right:
      Map of dust emission, extrapolated from IRAS/DIRBE maps to 410
      GHz. Note that the difference in the temperature range between the
      total and dust emission is a factor of $\sim40$.}
    \label{fig:dustmaps410}
  \end{center}
\end{figure*}

In Figures \ref{fig:dustmaps150}, \ref{fig:dustmaps240}, and
\ref{fig:dustmaps410} we show the MAXIMA-1 data as well as the SFD
extrapolated dust maps (using their preferred Model 8) at each of these
frequencies. Note that the temperature scale for the dust at 150 and 240
GHz is stretched considerably compared to the data -- the expected RMS
dust contribution is $\sim 1 \mu$K at 150 GHz, compared to the $\sim300\mu$K
(signal plus noise) RMS of the 150 GHz CMB map.

Because of this large difference in the amplitude of CMB and dust
emission, we do not expect to be able to see the dust contribution to a
single pixel in the maps. However, our procedure for estimating $\beta$
gives us the usual $(N_{\rm DOF})^{1/2}$ advantage when considering the
whole map ($N_{\rm DOF}$ gives the number of degrees of freedom in the
map, equal to the number of pixels less any degrees of freedom
marginalized over in making the map or by the methods described here).
We can  further increase the signal-to-noise by combining the individual
photometers at a given frequency. This allows us to compare the
different models offered by SFD, as detailed in their work.

\begin{figure}[htbp]
    \includegraphics[width=1.0\columnwidth]{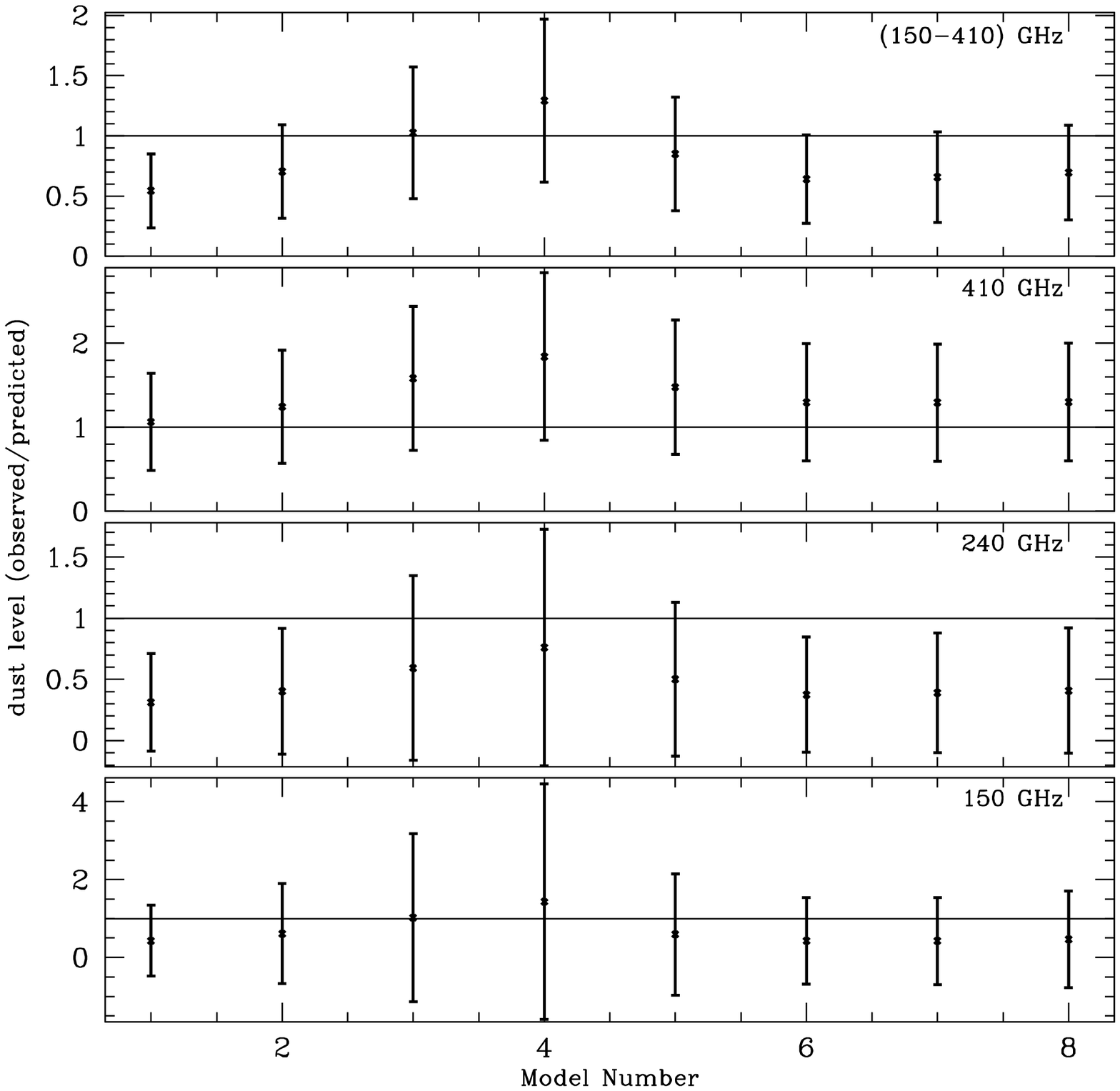}
  \caption{Each panel gives the observed amplitude of the dust template
    for the model number given on the horizontal axis at a given
    frequency. The top panel averages over all frequencies.} 
  \label{fig:allmods}
\end{figure}

In Figure~\ref{fig:allmods}, we show the observed amplitude $\beta$ for
each SFD model, for the MAXIMA-I detectors combined at each of 150, 240
and 410 GHz, and combined over all frequencies (the latter makes sense only
if the overall spectral shape of the model is correct over this
frequency range). The data do not strongly prefer any single
model. However, a few results are evident.

The dust signal is not strongly detected at 150 or 410 GHz for
\emph{any} model. That is, at these frequencies an amplitude of
$\beta=0$ is not disfavored. Indeed, some of the models are disfavored
at roughly the one sigma level from the 240 GHz data, with the
one-component model 1 the most disfavored.  At 410 GHz, the dust signal
is stronger, and $\beta=1$ (the model prediction) is preferred over
$\beta=0$ in all cases.

Despite the only marginal preference for a non-zero signal, these
results do contain important information. Even in the cases where
$\beta=0$ is allowed, the results can be interpreted as an upper limit
on the dust amplitude at these frequencies and in this area of sky. As
is evident from equations~\ref{eq:betamean} and \ref{eq:betavar}, the
observed amplitude and error scale inversely with the template
amplitude. Thus, although $\beta=1$ is acceptable for all of these
models, models predicting dust emission a factor of a few higher at any
of these frequencies would be strongly disfavored.

\begin{figure}[htbp]
  \centering
  \includegraphics[width=1.0\columnwidth]{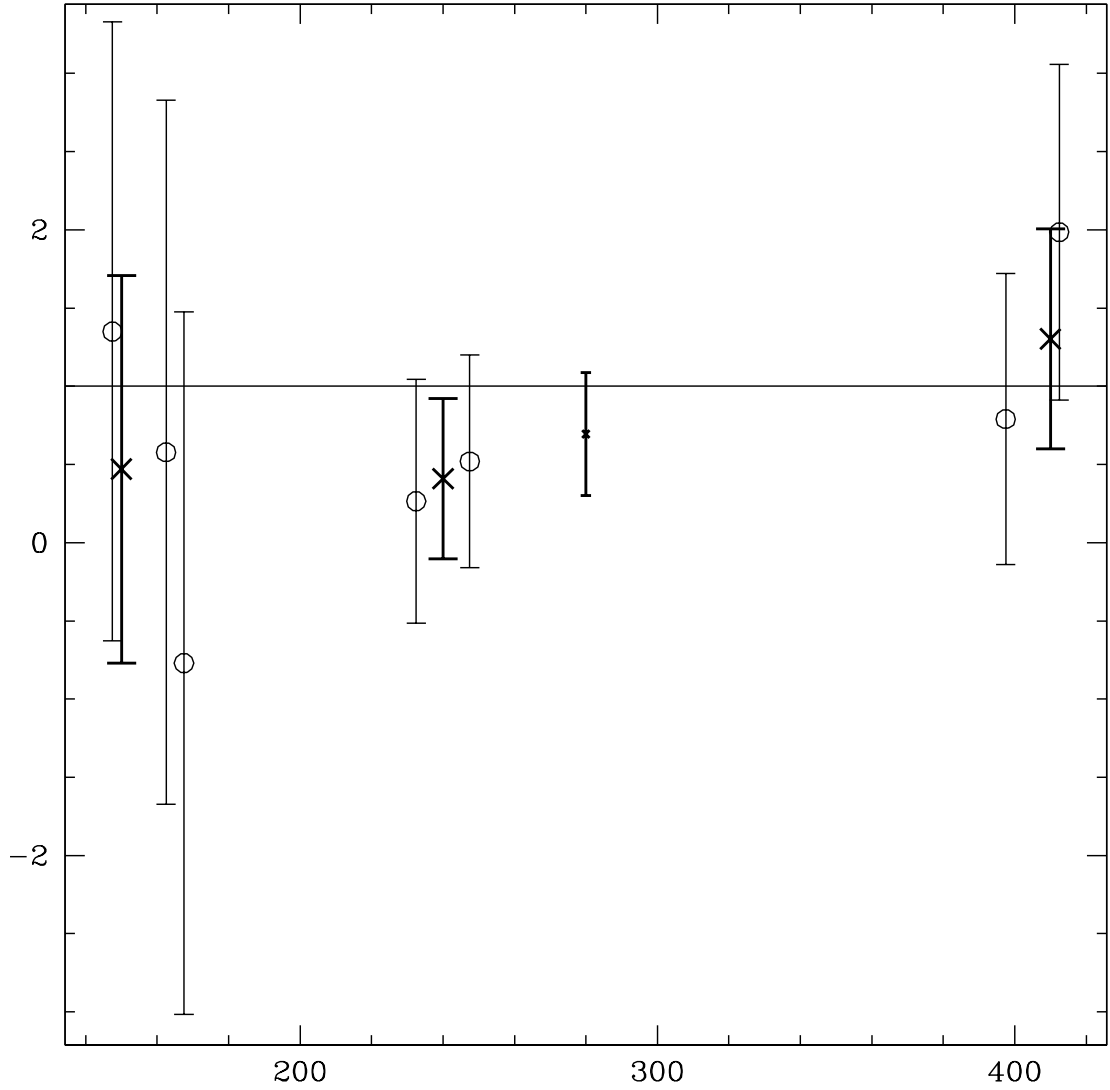}
  \caption{Ratio of expected amplitude of dust emission (SFD Model 8) to
    that observed in the MAXIMA-I maps, as a function of detector
    frequency (offset slightly for legibility). Thin error bars with circles
    are individual detectors, thick error bars with crosses at the
    individual frequencies (150, 240, 410) combine these, and the single
    error bar at $f\sim300$ gives the overall average.}
  \label{fig:amplitude}
\end{figure}
\begin{figure}[htbp]
  \centering
      \includegraphics[width=1.0\columnwidth]{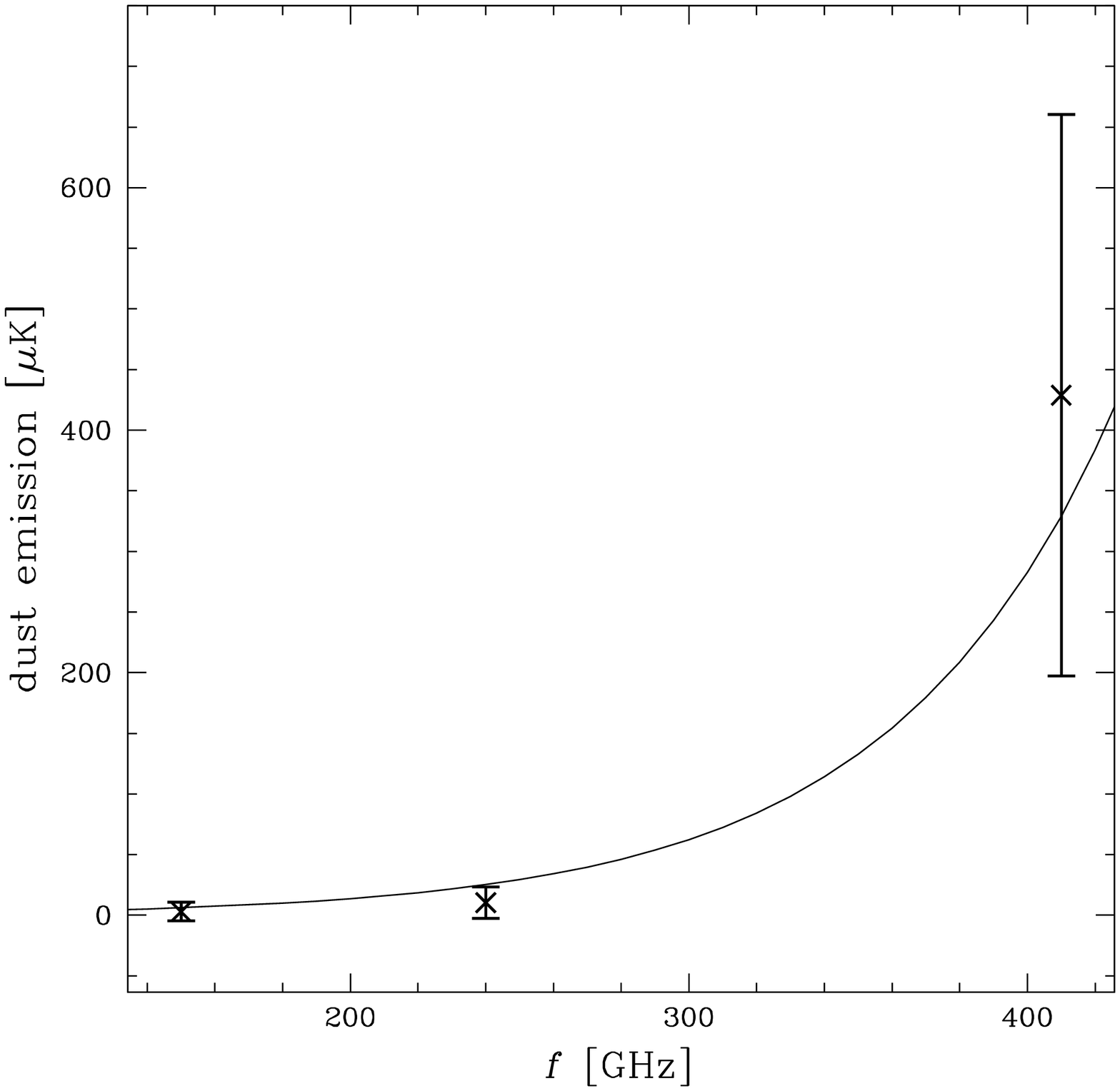}
  \caption{Thermodynamic temperature of dust emission observed in each of the three MAXIMA-I frequencies. The curve gives the average emission from Finkbeiner et al's Model 8 predicted in the MAXIMA-I observation area.}
  \label{fig:spectrum}
\end{figure}

In Figure~\ref{fig:amplitude}, we show the results for SFD's Model 8
--- their overall best fit --- in detail. We show each of the
individual detector amplitudes, the frequency averages, as well as an
overall average. In Figure~\ref{fig:spectrum}, we rescale the observed
coefficients $\beta$ to give the actual values of the observed dust
emission amplitude in thermodynamic temperature units.

This figure emphasizes that the techniques described here are not only
useful for the removal of CMB foregrounds, but are more generally useful
for estimating the foreground spectrum and extrapolating it to regimes
where it may be completely negligible in an individual pixel. 

\subsection{Synchrotron Emission}

We can of course use the same algorithm with other sources of foreground
emission for which we have maps.
The FSD code also provides an extrapolation synchrotron
emission as measured by \citet{Haslam,ReichReich,Rhodes}.

These surveys were reprocessed, destriped, and point-source subtracted
(D.P. Finkbeiner 2002, private communication) and are available to the public as part of
the dust map distribution\footnote{\url{http://astro.berkeley.edu/dust}}.

\begin{figure}[htbp]
  \centering
    \includegraphics[width=1.0\columnwidth]{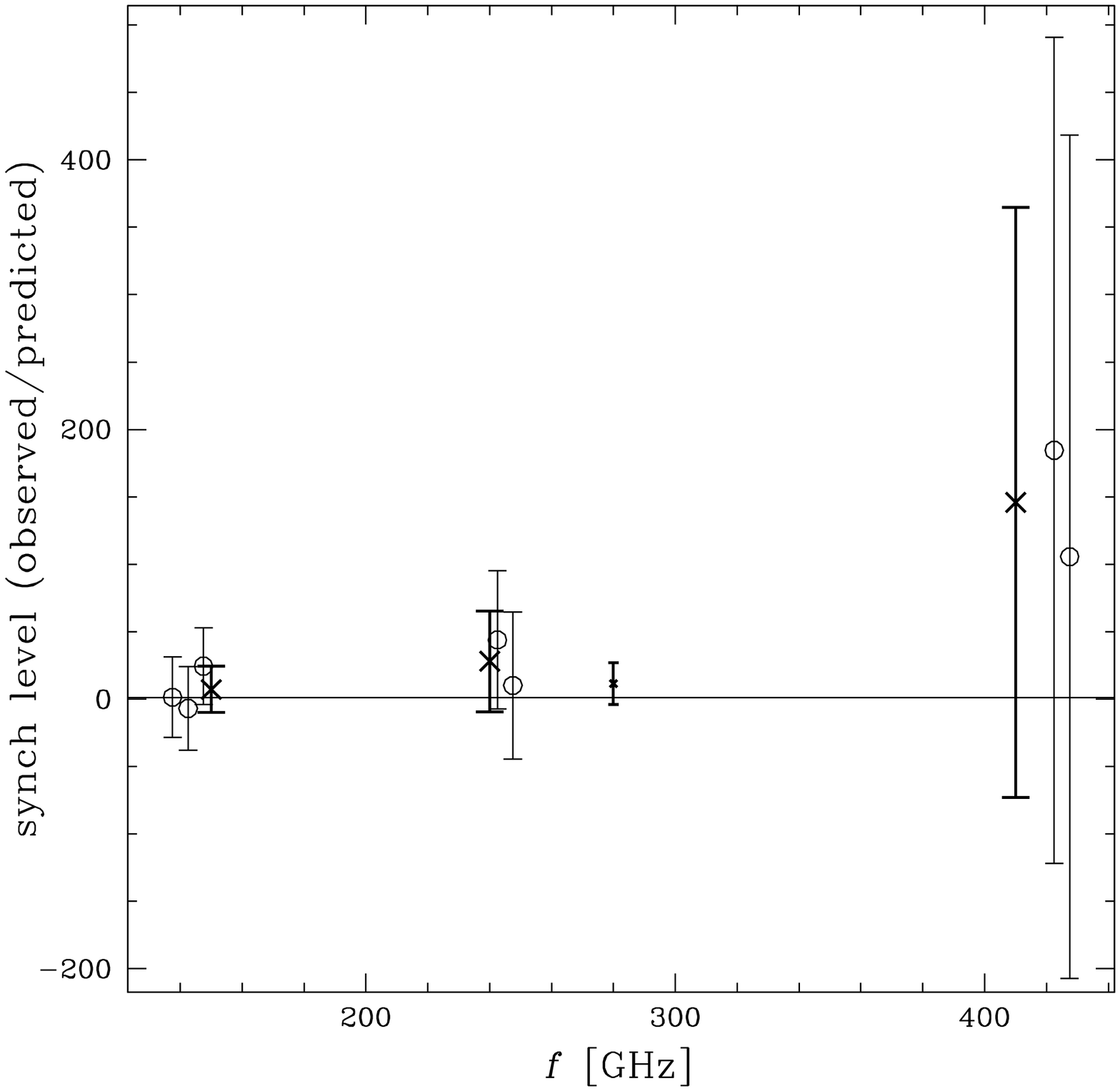}
  \caption{Observed synchrotron emission relative to that expected, as
    in figure~\ref{fig:amplitude}.}
  \label{fig:synch}
\end{figure}

The Haslam survey is full-sky, the Reich \& Reich survey is in the north, and
the Rhodes survey in the south, so for every point on the sky at least two
frequencies are available.  The surveys are beam-matched (at a one
degree resolution) and used to determine a power law for each pixel on
the sky.  This power law is then extrapolated to the frequency of
observation.  Because the synchrotron spectrum is known to fall faster
than a power law at high frequencies, this prediction should be
interpreted as an upper limit on the synchrotron emission.  If the 408
-- 2326 MHz maps are contaminated by significant free-free emission, the
power law slope is shallower than it should be, making it even more of
an upper limit.  The fact that even this upper limit is smaller than
$3\mu$K at 150 GHz over the MAXIMA area implies that the synchrotron
emission should be totally undetectable.  Figure~\ref{fig:synch}
confirms this to the extent possible with this data.


\subsection{The CMB power spectrum}

\begin{figure}[htbp]
  \centering
  \includegraphics[width=1.0\columnwidth]{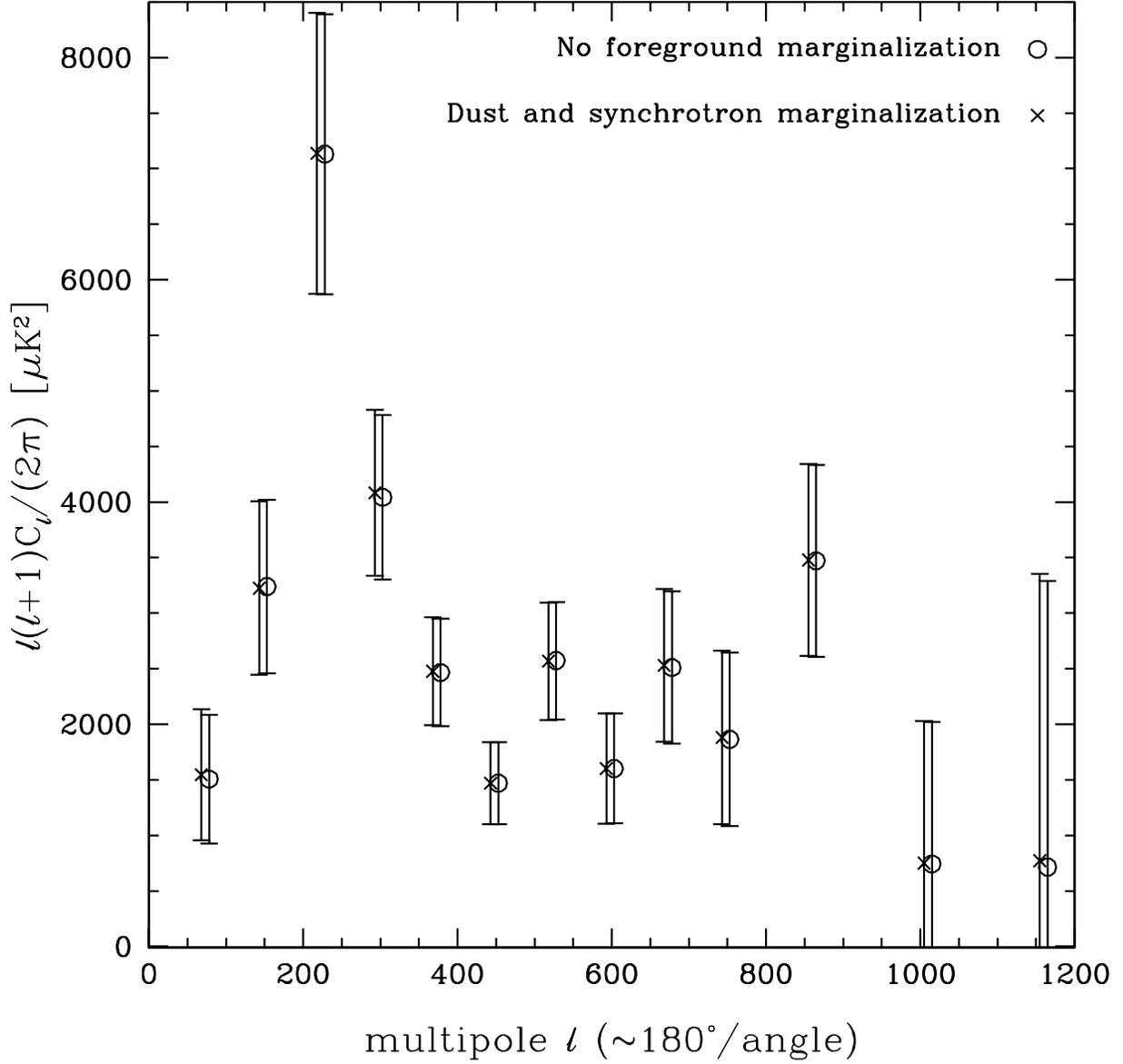}
  \caption{CMB Power spectrum ignoring the effect of dust and
    synchrotron contamination (left crosses) and marginalizing over
    it (right circles).}
  \label{fig:dustCl}
\end{figure}

Now that we have measured the overall level of dust emission in the
MAXIMA-I field, we can now ask the other questions posed in
Section~\ref{sec:methods}: what does the CMB itself look like?
In figure~\ref{fig:dustCl}, we show the power spectrum of CMB
temperature fluctuations from the combined 150 GHz photometers (as used
in \citet{Lee01} and \citet{Stompor01}). One set of points shows the
spectrum ignoring the contamination from dust and synchrotron emission
($\sigma_\beta\to0$ in equations~\ref{eq:Cllike} and \ref{eq:dmatrix});
the other set marginalizes over the dust emission with the known
morphology of FSD's model 8 ($\sigma_\beta\to\infty)$. We see that the
marginalization has very little effect --- much smaller than the error
bars. This is consistent with our knowledge of
the power spectrum of high-latitude dust emission, with
$C_\ell\sim\ell^{-3}$ (or perhaps closer to $\ell^{-2}$ in some parts of
the sky), dominating only at the largest scales.  This implies that the
spatial pattern of the dust (at least in the MAXIMA-I patch) is
essentially incompatible with that of an isotropic Gaussian field on the
sky.

\section{Discussion and future applications}

We have derived a technique for measuring and accounting for the effect
of foreground emission on CMB observations, for the case where the
morphology of the contaminant is known, but when its spectrum is unknown
or imprecisely measured. We have applied these techniques to the
MAXIMA-1 data and observations of dust and synchrotron emission. The
dust emission in the MAXIMA-I region is consistent with models and
observations at higher frequencies, although it has negligible effect on
the measured CMB power spectrum.

The MAXIMA-1 observing region was specifically chosen to be a region of
low dust contrast. As high-resolution CMB observations cover more of the
sky with higher sensitivity, these techniques will become more important
for the separation of the various components. 

These techniques have many further applications, some alluded to above.
We can allow for spatial variation in the foreground spectrum and/or
inaccuracies in our foreground templates. Most straightforwardly, we can
apply the technique separately to individual patches (with, say,
different dust temperatures) and allow the foreground amplitudes to
float separately between them. Such a technique could be applied
iteratively on smaller patches until the signal-to-noise of the result
decreased too far. In particular, we would certainly split the full sky up into regions within and outside of the Galactic plane, where we know the foreground properties to differ.
Other prior knowledge will affect the
algorithm in different ways. If we thought that the foreground spectrum
was approximately a power law, $f_p\propto\nu^b$, we can estimate (or
marginalize over) the power law index. The change in the
foreground amplitude is $f_p\delta b\ln\nu$, equivalent to the $\beta
f_p$ in Eq.~\ref{eq:data}.

More ambitiously, if we have knowledge of foreground
\emph{polarization}, these techniques carry forward identically,
although such measurements may not be readily forthcoming.

Finally, we have discussed here the case where the foreground template
is known with considerably more accuracy than the CMB
measurement. As CMB observations are performed with higher sensitivity,
we will need to deal with foreground templates with errors of their own.
Although the calculations and resulting algorithms are
considerably more complicated, the same general setting can be used; the
results are similar to the cases discussed in \cite{Knoxetal1998}.

\acknowledgments

We thank Danny Ball and the other staff at NASA's National Scientific
Balloon Facility in Palestine, TX for their outstanding support of the
MAXIMA program.  MAXIMA is supported by NASA Grants NAG5-3941,
NAG5-6552, NAG5-4454, GSRP-031, and GSRP-032, and by the NSF through the
Center for Particle Astrophysics at UC Berkeley, NSF cooperative
agreement AST-9120005, and a KDI grant 9872979.  The data analysis used
resources of the National Energy Research Scientific Computing center
which is supported by the Office of Science of the U.S. Department of
Energy under contract no.\ DE-AC03-76SF00098. RS acknowledges support
from NASA's grant S-92548-F. PGF and AHJ acknowledge
support from PPARC in the UK.


\end{document}